# Fermi Communications and Public Outreach

L. Cominsky, A. Simonnet and the Fermi E/PO team

*Sonoma State University, Rohnert Park, CA, 94928 USA*

The Sonoma State University (SSU) Education and Public Outreach (E/PO) group participates in the planning and execution of press conferences that feature noteworthy *Fermi* discoveries, as well as supporting social media and outreach websites. We have also created many scientific illustrations for the media, tools for amateur astronomers for use at star parties, and have given numerous public talks about *Fermi* discoveries.

## 1. PRESS CONFERENCES AND PRESS RELEASES

Beginning with the activities leading up to the launch of *Fermi* on June 11, 2008, there have been many press conferences, media telecons, press releases and news features that showcase the discoveries and news about *Fermi*. Table I summarizes the number of news releases and features issued each year since launch, as well as each year's top stories that were showcased in press briefings or media telecons. Many of the press briefings occurred at scientific conferences including: the American Astronomical Society (AAS) winter and summer meetings, the American Physics Society (APS) April meeting, the AAS High Energy Astrophysics Division (HEAD) meeting, and the American Geophysical Union (AGU) annual meeting. For complete links see
   http://fermi.gsfc.nasa.gov/ssc/library/news/

### 1.1. Science Magazine Covers

*Fermi* discoveries in 2009 and 2014 were the subject of cover stories in *Science* magazine. Illustrations by Aurore Simonnet that depicted the discoveries were chosen for cover art. The 2009 publications featured *Fermi* observations of pulsars while the 2014 cover portrayed GRB 130427A, one of the brightest gamma-ray bursts ever seen. This GRB was observed by many experiments on Earth and in space, including *Fermi*. Figure 1 shows the *Science* cover art illustrating GRB 130427A.

### 1.2. Fermi Pulsar Interactive Explorer

Developed by SSU's Kevin John, this interactive map illustrated the Fermi pulsar discoveries that were highlighted in a media telecon on 3 November 2011. For each pulsar observed by Fermi, the interactive provides the pulse rate, location and a user-friendly description of the pulsar's significant observations. The pulsar videos in this interactive have been viewed more than 145,000 times to date. The pulsar interactive can be viewed at:
   http://www.nasa.gov/externalflash/fermipulsar/

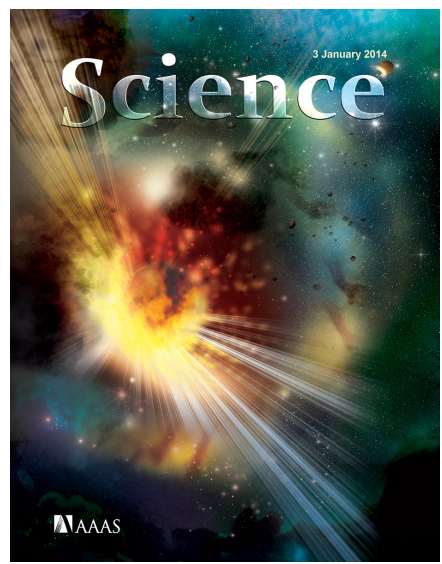

Figure 1: *Science* magazine cover from 3 January 2014 illustrating GRB 130427A, a "shockingly bright" gamma-ray burst

### 1.3 YouTube and SVS Videos

NASA Goddard's Scientific Visualization Studio (SVS) employs many talented animators and illustrators that help explain *Fermi*'s scientific discoveries to the public. Many press briefings, media telecons, press releases and news features include illustrations and video products that help to explain the extreme Universe that *Fermi* observations are revealing. Since launch, the number of views of *Fermi* videos on the SVS website (http://svs.gsfc.nasa.gov) has exceeded 3.4 million. Table II lists *Fermi* SVS videos with more than 100,000 views, along with viewing statistics and the SVS reference numbers.

Beginning in 2011, animated media products that were created to illustrate press releases and briefings have been uploaded to NASA's YouTube channel. Table III lists *Fermi* YouTube videos with more than 50,000 views, along with viewing statistics and the YouTube reference codes.

Table I Media Releases and Briefings

| Year | Releases | Media Events |
|---|---|---|
| 2008 | 9 | Launch Coverage, First Light |
| 2009 | 13 | Pulsars (AAS), Space Time (NASA), Star Factories (Fermi Symposium) |
| 2010 | 7 | Millisecond Pulsars (AAS), SNR (APS), Extragalactic Background (HEAD), Fermi Bubbles (NASA) |
| 2011 | 9 | Anti-matter from Thunderstorms, Non-constant Crab (AAS), Millisecond Pulsar (NASA) |
| 2012 | 5 | High-energy Sky (AAS), Dwarf Galaxies and Dark Matter (APS), Ancient starlight (Fermi Symposium) |
| 2013 | 8 | Black Hole Flares (AAS), Fermi Turns 5 (NASA) |
| 2014 | 9 | Gravitational Lens (AAS), Blazar Batteries (AAS), Thunderstorms (AGU) |

Table II Fermi SVS Videos with more than 100,000 views to date

| Video | Date | Views | SVS Number |
|---|---|---|---|
| GLASTcast for iTunes | 3 Jun 2008 | 321,408 | 10250 |
| Einstein's Cosmic Speed Limit | 28 Oct 2009 | 258,721 | 10510 |
| Terrestrial Gamma-ray Flashes Create Antimatter | 10 Jan 2011 | 169,945 | 10706 |
| Fermi Discovers Youngest Millisecond Pulsar | 3 Nov 2011 | 160,499 | 10858 |
| Fermi Sees a Nova | 12 Aug 2010 | 150,080 | 20184 |
| Fermi Pulsar Interactive Videos | 3 Nov 2011 | 145,118 | 10861 |
| Fermi Observations of Dwarf Galaxies Provide New Insights on Dark Matter | 2 Apr 2012 | 117,882 | 10943 |
| Stellar Odd Couple Makes Striking Flares | 29 Jun 2011 | 109,434 | 10798 |
| Fermi's Latest Gamma-ray Census Highlights Cosmic Mysteries | 9 Sep 2011 | 105,726 | 10819 |

## 2. PUBLIC OUTREACH

From the top of Mt. Tamalpais to seniors in Oakmont to amateur astronomers all across the USA, E/PO lead Lynn Cominsky has given dozens of public lectures about blazing galaxies, monstrous black holes and the extreme Universe as seen by *Fermi*. Other US *Fermi* team members who have given many public talks include LAT Principal Investigator Peter Michelson, Project Scientist Julie McEnery, Deputy Project Scientist David Thompson, and team member Roopesh Ojha.

The "make your own pulsar model" activity is one of *Fermi*'s most popular public engagements, and was originally featured on the back of the *Fermi* lithograph and in the Supernova Educator's Guide, both developed by the SSU E/PO team. This shining model is suitable for kids of all ages, and teaches about pulsars as well as about simple circuit design, using a battery and an LED. It has been showcased at the American Astronomical Society student engagement events for the past three years, as well as at many public open house events, including the SLAC-KIPAC open house, school science fairs, and the North Bay Discovery Days. Figure 2 shows Lucy and Abby Dilbeck demonstrating the pulsar model at a recent SLAC-KIPAC Open House.

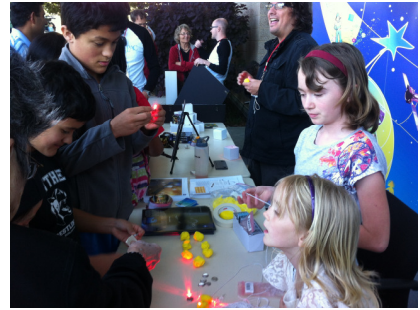

Figure 2: Lucy and Abby Dilbeck demonstrate how to make your own model pulsar

### 2.1. Epo's Chronicles

From 2008 - 2013, the SSU E/PO team produced over 200 weekly "eposodes" of Epo's Chronicles, a web comic that illustrated the adventures of Alkina and her sentient spaceship Epo. Alkina and Epo traveled through the galaxy, learning about space science and searching for their origins. Translated from English into French, Italian and Spanish, these popular comic strips were viewed by thousands each month. During 2012 (the last complete year of the webcomic), over 80,000 unique IP addresses viewed the site. External evaluations of Epo's Chronicles indicated that "Participants particularly liked the "Web 2.0" aspect of the comic, and the use of links to learn more and pursue various topics in a multimedia platform." In addition, the "artwork was highly praised." Figure 3 shows one of the comics related to *Fermi*.

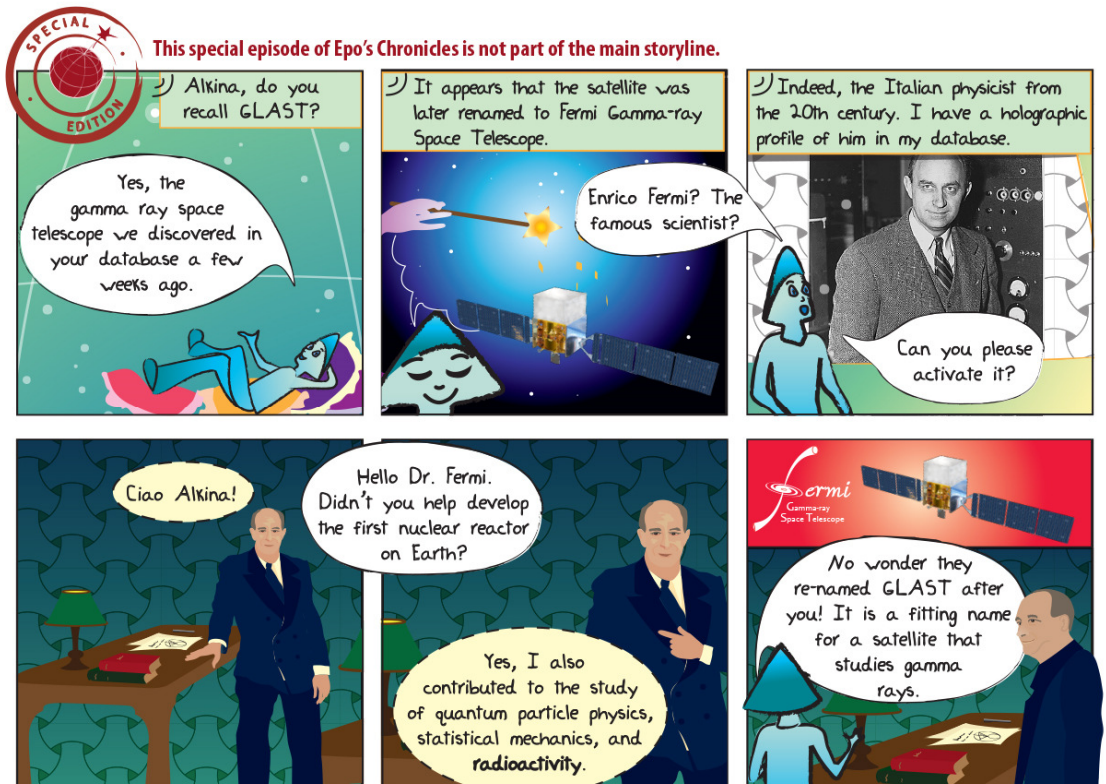

Figure 3: Epo's Chronicles special "eposode" about *Fermi*

## 2.2. Citizen Science through a *Fermi*-LIGO Collaboration

The collaboration between *Fermi* and the Laser Interferometer Gravitational-wave Observatory's Einstein@Home project was advertised with a postcard-sized handout inviting participants to try to discover a gamma-ray pulsar using *Fermi* data. Several have been discovered by citizen scientists running the Einstein@Home program on their home computers as a screen saver. The postcard is shown in Figure 4.

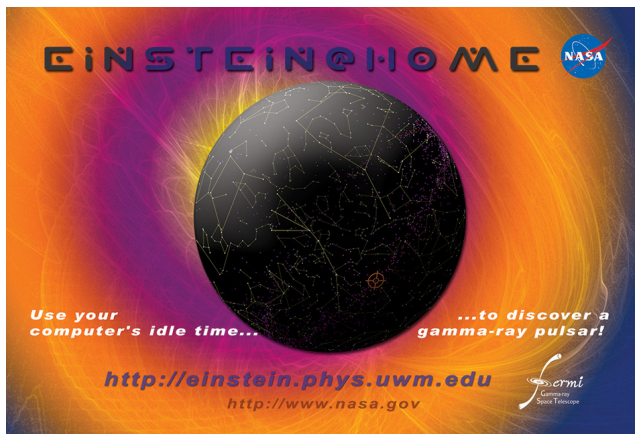

Figure 4: Einstein@Home postcard inviting the public to search for pulsars in the *Fermi* data

## 2.3. *Fermi* Skymap Poster

The poster shown in Figure 5 was created in 2012 by Aurore Simonnet for distribution at the *Fermi* Symposium in Rome. Eight major *Fermi* discoveries are called out from the iconic image of the high-energy gamma-ray sky as seen by the Large Area Telescope through 2011. The discoveries that are illustrated include:

- CTA1, the first gamma-ray-only pulsar
- Nova V407 Cygni, the first gamma-ray nova
- Repeated gamma-ray flares from the active galaxy 3C454.3
- Resolved GeV gamma rays from the supernova remnant W44
- Giant gamma-ray lobes emanating from the center of the Milky Way now known as the *Fermi* bubbles
- Resolved extended gamma rays surrounding the active galaxy Centaurus A
- Flaring gamma-ray emission and changing x–ray emission from the Crab nebula, previously thought to be a constant "standard candle"
- GRB 090510A, the distant short gamma-ray burst that was used to set limits on the foaminess of spacetime

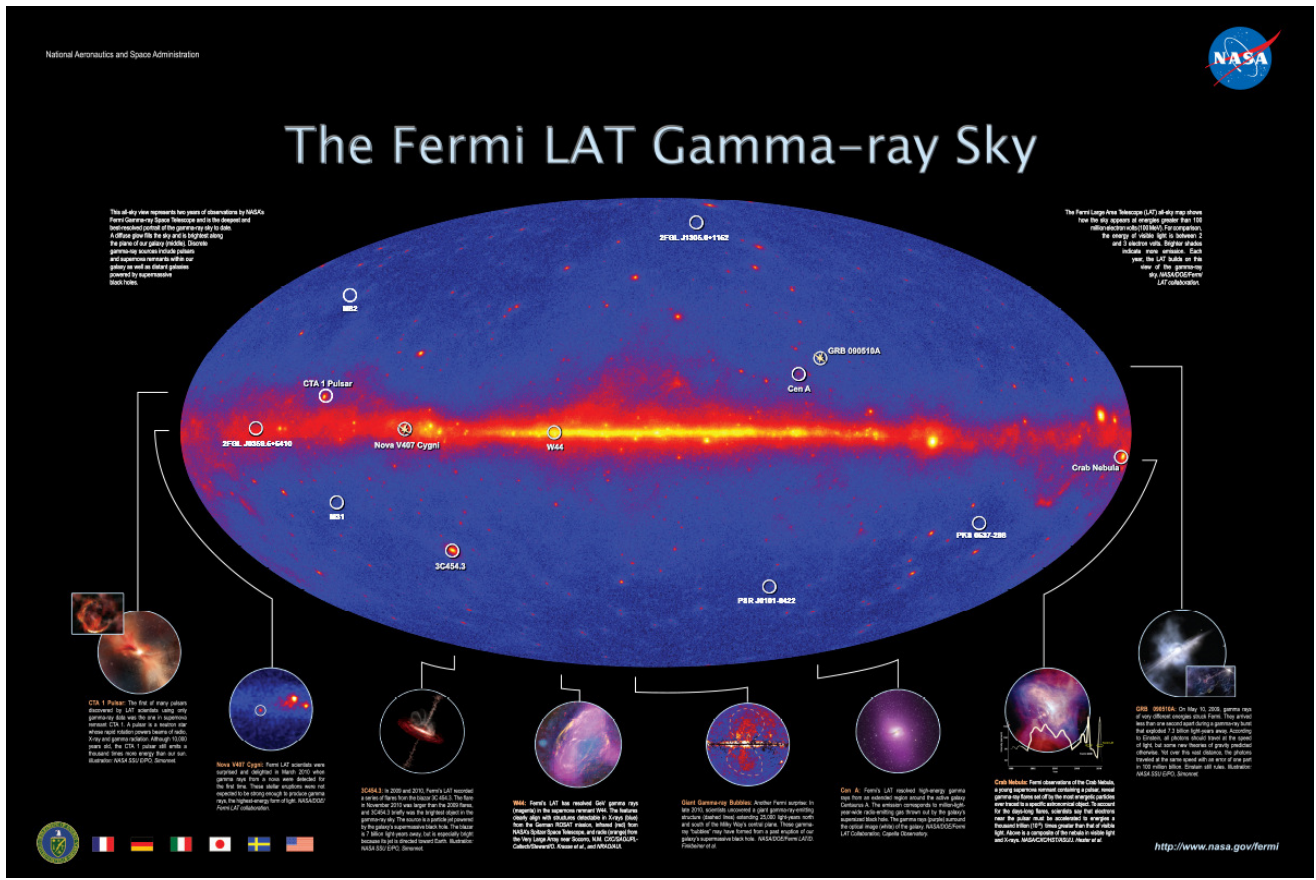

Figure 5: Two-year *Fermi* skymap with highlights printed in 2012

### 2.4. Amateur Astronomers

Public outreach is often conducted by amateur astronomers through star parties held nation-wide. The *Fermi* E/PO team co-sponsored the SUPERNOVA! toolkit for use by these passionate advocates for astronomy. Since 2008, when the toolkit went into national distribution, it has reached over 138,000 attendees through more than 1300 events. Of these events, 680 events reported including almost 25,000 minorities and over 39,000 women/girls.

### 2.5. *Fermi* Exhibits

The *Fermi* exhibit booth has undergone many transformations over the years. The most recent booth graphics feature the *Fermi* skymap silhouetted with an image of the satellite as shown in Figure 6. The exhibit booth is often accompanied by the *Fermi* banner stand, which features a blueprint style graphic that illustrates the project logo, the satellite, the flags of the participating countries, and the skymap. The exhibit booth and/or banner have been used at venues including the AAS winter meetings, the USA Science and Engineering Festival, and the Goddard Jamboree.

The SSU E/PO team has a multi-mission exhibit booth drawn in the style of Epo's Chronicles that includes images of Alkina and other characters from the web comic. This booth has been used extensively at educator and student events, including California Science Teachers Association annual meetings, Expanding Your Horizons, SSU Seawolf Day, and the North Bay Science Festival.

### 2.6. Social Media

*Fermi*'s presence in the world of social media includes a Facebook page and the Twitter feed @NASAFermi. Since launch, there have been 300 tweets from @NASAFermi, and the feed has over 35,000 followers. The *Fermi* Facebook page has over 30,000 likes and can be found at: `http://www.facebook.com/nasafermi`.

### 2.7. International Year of Astronomy

The year 2009 was the International Year of Astronomy (IYA). Public outreach events occurred throughout the world, and the *Fermi* E/PO team participated in many of them, including the creation of special illustrated lithographs featuring the objects of the month as explained by Alkina from the Epo's Chronicles webcomic. Over 18,000 of these lithographs were distributed nationwide through amateur astronomy clubs through NASA's Night Sky Network.

Another special creation was an Epo's Chronicles podcast distributed through the 365 Days of Astronomy website. This podcast was downloaded over 6000 times following its initial release on 16 September 2009.

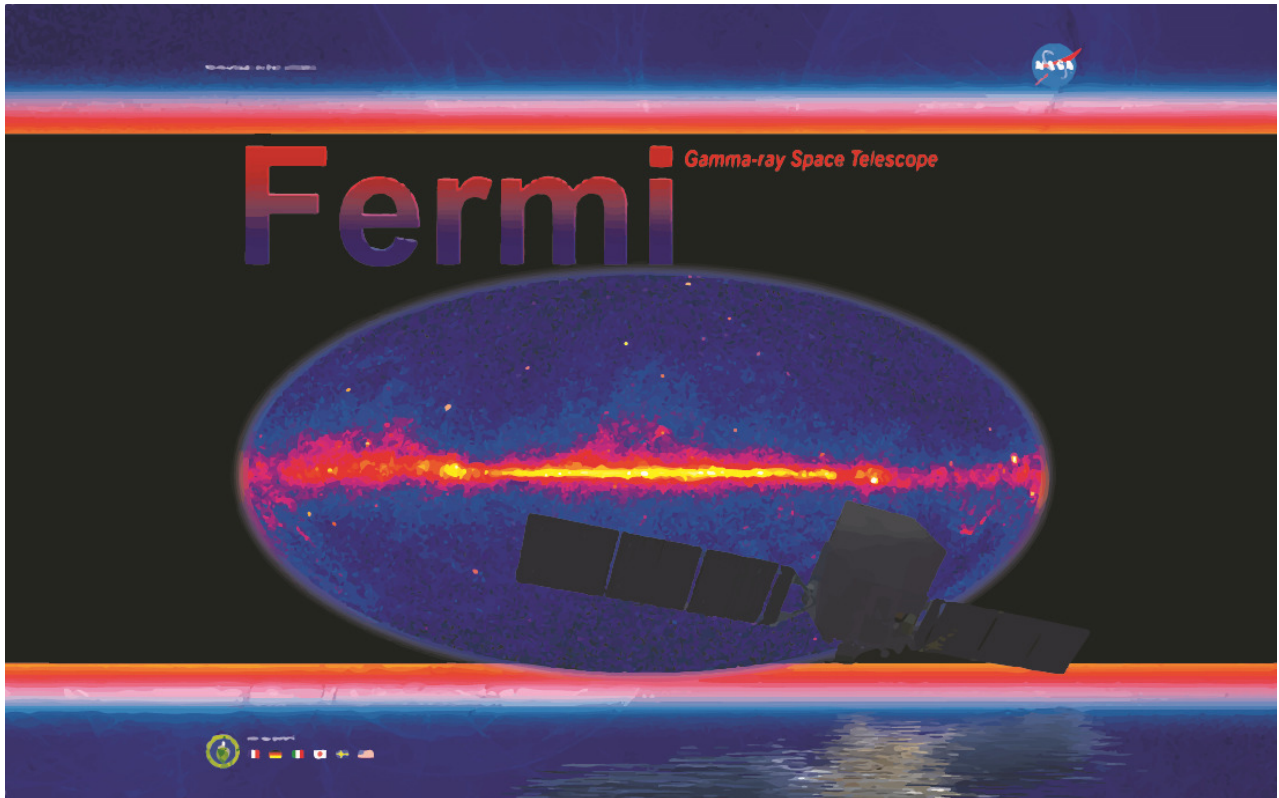

Figure 6: Current *Fermi* exhibit booth graphics

SSU E/PO created a traveling exhibit of IYA images that circulated around the San Francisco Bay Area during 2009. This small exhibit was featured at 20 venues, with an estimated viewing by over 100,000 participants. A larger IYA exhibit appeared at the California Academy of Sciences and San Jose Tech Museum, with estimated viewing of 50,000 at each location.

### 2.8. Black Hole Shows

"Black Holes: The Other Side of Infinity" and the PBS NOVA show "Monster of the Milky Way" were produced in partnership with *Swift*, the National Science Foundation, the Denver Museum of Nature & Science, PBS NOVA and Tom Lucas Productions. Premiering in 2006, the planetarium show has been featured in over 30 venues and has reached millions of people world-wide. Narrated by Liam Neeson, this full-format digital dome show included state-of-the art scientific visualizations of black holes and warped spacetime created by experts at the National Center for Supercomputer Applications at the University of Illinois Urbana-Champaign. The PBS NOVA show was initially seen by over 10 million viewers and has aired many times since then. It is still available for viewing on the PBS website. The black hole shows were initially seed-funded by *Fermi* E/PO and Lynn Cominsky served as a scientific director.

### 2.9. Printed Materials

Many printed materials were developed by the *Fermi* E/PO team for distribution to the general public. Table IV summarizes the number of these items that were handed out during 2000 - 2013. Prior to the renaming of the mission in late 2008, these products listed the satellite name as GLAST (Gamma-ray Large Area Space Telescope) rather than *Fermi*.

The *Fermi* sticker features a colorful image of the satellite on the front along with text describing the mission on the back. It is shown in Figure 7.

The *Fermi* lithograph features an illustration of the satellite on the front, and an explanation of the overall scientific objectives of the mission on the back. Instructions for the "make your own pulsar" activity are also included.

The *Fermi* fact sheet is a four-page color brochure that describes the science of the mission, as well as providing tables that summarize the instrumental parameters and the mission participants.

The *Fermi* brochure describes in detail the science that *Fermi* does and explains how it does it. The description includes the instruments, background information on gamma-ray astronomy and detection methods. It also describes pre-launch thinking about about active galaxies, gamma-ray bursts, solar flares, gamma-rays from dark matter and other highly energetic sources seen in the Universe.

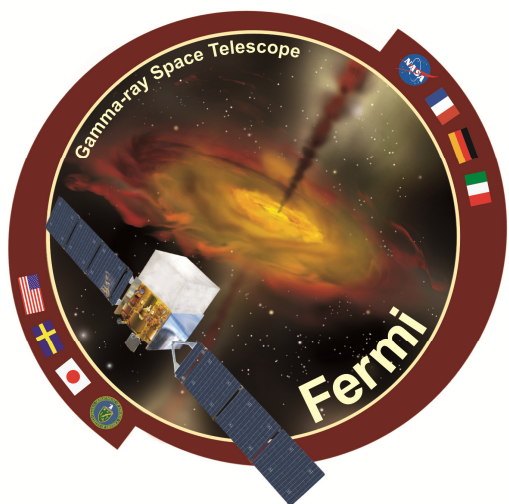

Figure 7: Official Fermi mission sticker

The *Fermi* paper model provides a short description of the scientific instruments on board the satellite, as well as links to other resources about its instruments. There is also a short description of how *Fermi* detects gamma rays with the Large Area Telescope as well as the Gamma-ray Burst Monitor detectors. The product includes three pages of parts that can be cut out and easily assembled using common household items.

Table IV  Printed Materials

| Product | Created | Number Distributed |
|---|---|---|
| Stickers | 2005 | 45,500 |
| Lithograph | 2008 | 10,000 |
| Fact Sheet | 2008 | 9500 |
| Brochure | 2008 | 15,000 |
| Paper model | 2007 | 8800 |
| Race card game | 2005 | 7350 |
| Black Hole FAQ | 2006 | 32,000 |

The *Fermi* Race Card game challenges two teams of players to strategically maneuver to be the first to assemble the parts of the satellite and then use it to observe five astronomical objects. As players build their satellites they must overcome hurtles and obstacles thrown at them by their opponents while doing the same in order to slow their opponents down. To win, players must successfully have their operational *Fermi* satellite observe five gamma-ray emitting objects.

The black hole frequently-asked questions (FAQ) brochures answers eight of the most commonly asked questions about black holes, and explains how *Fermi* studies black holes. The FAQ brochures were distributed to attendees at many of the planetaria who experienced "Black Holes: the Other Side of Infinity."

### 2.10. Tesla Coil Show

From 2000-2012, *Fermi* E/PO provided funding to support the Tesla Coil show put on by scientists and students from the University of California, Santa Cruz Institute for Particle Physics. These popular shows reached thousands of students annually.

### Acknowledgments

The SSU Education and Public Outreach group would like to acknowledge support from NASA NNX07AF53G, NNX12AE34G and NAS5-00147 through Stanford University. We also would like to thank the many talented science writers and animators at NASA Goddard Space Flight Center with whom we have worked during the years including Christopher Wanjek, Robert Naeye, Francis Reddy, Scott Wiessinger, and the staff at Goddard's Scientific Visualization Studio.